\documentclass[10pt, conference]{IEEEtran}

\usepackage{comment}
\usepackage[table,xcdraw]{xcolor}
\usepackage{cite,url}
\usepackage[textsize=tiny]{todonotes}
\usepackage{graphicx, tikz,float,caption}
\usepackage{dblfloatfix,listings,courier}
\usepackage{xcolor,multirow,array,xspace}
\usepackage[textsize=tiny]{todonotes}
\usepackage{multicol}

\begin{document}
%

\title{Stack Overflow in Github: Any Snippets There?}


\author{\IEEEauthorblockN{Di Yang, Pedro Martins, Vaibhav Saini and Cristina Lopes}
\IEEEauthorblockA{Department of Informatics\\
University of California, Irvine\\
Irvine, USA\\
Email: \{diy4, pribeiro, vpsaini, lopes\}@uci.edu}
}


%


\maketitle

\begin{abstract}
When programmers look for how to achieve certain programming tasks,
Stack Overflow is a popular destination in search engine results. Over
the years, Stack Overflow has accumulated an impressive knowledge base
of snippets of code that are amply documented. We are interested in
studying how programmers use these snippets of code in their
projects. Can we find Stack Overflow snippets in real projects? When
snippets are used, is this copy literal or does it suffer adaptations?
And are these adaptations specializations required by the
idiosyncrasies of the target artifact, or are they motivated by
specific requirements of the programmer? The large-scale study
presented on this paper analyzes 909k non-fork Python projects hosted
on Github, which contain 290M function definitions, and 1.9M Python snippets
captured in Stack Overflow. Results are presented as quantitative
analysis of block-level code cloning intra and inter Stack Overflow
and GitHub, and as an analysis of programming behaviors through the
qualitative analysis of our findings.
\end{abstract}


\IEEEpeerreviewmaketitle

\section{Introduction}

The popularity and relevance of the Question and Answer site Stack
Overflow (SO) is well known within the programming community. As a
measure of its populatiry, SO received more than half a billion views
on the first 30 days of 2017
alone\footnote{\url{https://www.quantcast.com/stackoverflow.com}
  [Accessed January, 2017]}. Another very popular site is Github (GH),
a project repository that ranked 14$^{th}$ on Forbes Cloud 100 in
2016\footnote{The Forbes Cloud 100 recognizes the top 100 private
  cloud companies in the world
  (\url{http://www.forbes.com/cloud100}).}  Although both sites are
equally relevant for the programming community, they are so in
different contexts. SO is a Q{\&}A website with a strong
community-based support, responsible for providing answers for
virtually any type of programming problems and helping any type of
user, from casual SHELL users to expert system administrators. GH also
has a strong social component, but it is more focused on the storage
and maintenance of software artifacts, providing version controlling
features, bug management, control over the coder-base and contributors
of projects, and so on.

Both platforms are part of a larger system of globalized software
production. The same users that rely on the hosting and management
characteristics of GH often have difficulties and need help on the
implementation of their computer programs, seek support on SO for
their specific problems, or hints of solutions from ones with a degree
of similarity, and return to GH to apply the knowledge
acquired. Empirically, however, there is little evidence of the actual
impact that these two systems have on each other, or of the kind of
information that goes from one platform to the other. Analyzing this
relation is the focus of this work.

In isolation, SO has been the subject of various research
studies. One example is the use
of topic modeling on SO questions to categorize discussions
\cite{Bajaj:2014,Wang:2014,Arwan:2015}, another is the use SO
statistics to analyze use behavior and activity
\cite{Vasilescu:2014,Chen:2016}. Recent work has paid special
attention to code snippets. Wong \textit{et al.}~\cite{Wong:2013} and
Ponzanelli \textit{et al.}~\cite{Ponzanelli:MSR2014} both mine SO for
code snippets that are clones to a snippet in the client system. Yang
\textit{et al.}~\cite{Yang:2016} provide a usability study of code
snippets of four popular programming languages.

There are already some studies that investigate some relations between SO
and GH. Vasilescu \textit{et al.}~\cite{Vasilescu:2013} investigated
the interplay between asking and answering questions on SO and
committing changes to GH repositories. They answered the question of
whether participation in SO relates to the productivity of GH
developers. From this work, we know that GH and SO overlap in a
knowledge-sharing ecosystem: GH developers can ask for help on SO to
solve their own technical challenges; they can also engage in SO to
satisfy a demand for knowledge of others, perhaps less experienced
than themselves. Moreover, we see this overlapping of knowledge also
indicating another kind of overlapping: pieces of code. GH programmers
can copy-paste SO code snippets to solve their particular problems;
they can also use their existing code in GH repository to answer SO
questions.

An \textit{et al.}~\cite{An:2017} conducted a case study with 399 Android apps, to investigate whether developers respect license terms when reusing code from
SO posts (and the other way around). They found 232
code snippets in 62 Android apps that were
potentially reused from SO, and 1,226 SO
posts containing code examples that are clones of code released
in 68 Android apps, suggesting that developers may have copied
the code of these apps to answer SO questions.

In this study, our goal is to investigate and understand how much the
snippets obtained from SO are used in GH projects. We opertionalize
this problem as pieces of source code that exist in both sides, and we
search for cloning and repetition as a measure of equal information
presented in both places.

How much of the knowledge base, represented as source code,
is shared between SO and GH? If SO and GH have overlapping source
code, is this copy literal or does it suffer adaptations? And are
these adaptations, if they exist, specializations required by the
idiosyncrasies of the target or by the idiosyncrasies or the
programmer, or both? 

To answer these questions we perform intra and inter code duplication
analysis on GH and SO. We uncover and document code duplicates in 909k
Python projects from GH, which contain 292M function
definitions in GH and 1.9M snippets in SO. Our choice of language is
driven by popularity and by existing work by Yang \textit{et
  al.}~\cite{Yang:2016}, which shows Python snippets in SO having one
of the highest usability rates among the popular languages.

The rest of the paper is organized as
follows. Section~\ref{sec:methods} details the methodology we applied
to find code duplicates. Section~\ref{sec:dataset} describes the
datasets we used. Quantitative findings are presented in
Section~\ref{sec:findings-quant} and qualitative analysis in
Section~\ref{sec:findings-qual}. Related work is present in
Section~\ref{sec:related-work}. Section~\ref{sec:conclusion} concludes the paper.

\section{Methodology}
\label{sec:methods}

\begin{figure*}
\centering{
\includegraphics[scale=.5]{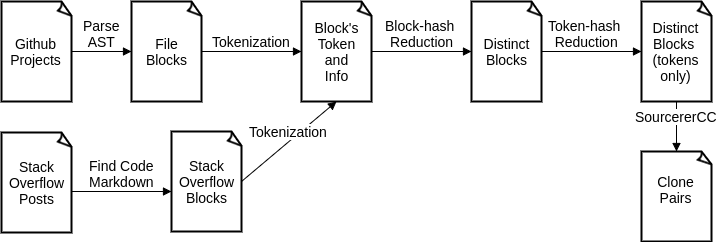}
}
\caption{Pipeline for file analysis.}
\label{fig:pipeline}
\end{figure*} 

In this section, we describe the pipeline followed for analyzing block-level duplication inter and intra GH and SO.

Figure \ref{fig:pipeline} contains the main steps in the analysis process. The pipeline starts by extracting blocks from GH projects and SO posts. Blocks from both origins are then scanned to obtain tokens and other relevant information (a process we call tokenization from now on).

In this analysis process, we provide three levels of similarity: a hash on the block, a hash on tokens, and an 80\% token similarity.
These capture the case where entire blocks are copied as-is, smaller changes are made in spacing or comments, and more meaningful edits are applied to the code.

Moreover, all the similarity analyses are done for intra-GH, intra-SO, and inter GH and SO.
The following of this section will discuss each of the step in the pipeline in detail.

\subsection{Extracting blocks from a Python file}
\label{subsec:parse}

For GH projects, our concept of block is that of a function
definition. We extract the following two kinds of
function definitions:
(1) function defined inside of a class; 
(2) function defined outside of a class;
For nested functions, we only consider the outtermost function.
Consider the following example:

\begin{lstlisting}[language=Python, label={lst:func-example},caption={Github blocks}]
class Foo:
  def func1(a, b, c):
    return a + b

def func2(a, b, c):
  if a>b:
    return c
  return 0
		
def func3(a):
  def func4(b):
    return b*2
  return func4(3)
\end{lstlisting}

From the example above, we extract three functions: \lstinline|func1|,
\lstinline|func2|, and \lstinline|func3|. \lstinline|func4| was nested inside of an already existing block, \lstinline|func3|, and
is therefore ignored.

The AST also exposes the starting and ending line numbers for its
constituents, information we use to define blocks and contextualize
them. Note that this is only possible in settings where a block
resides within a file, such as GH; for SO the line numbers are
useless.

\subsection{Extracting blocks from SO posts}
\label{subsec:multiline}

In SO, both questions and answers are considered Posts, for which a unique id is associated. Posts are distinguished by a \lstinline|PostTypeId| indicating if it is a question \lstinline|PostTypeId=1| or an answer \lstinline|PostTypeId=2|. The link between answers and their original questions is preserved. Only Question posts have tags marking the related languages and topics of the post, therefore all the pieces of code we process come from, or are related to, a Question whose tags contain 'python'. For all posts for Python, we used the markdown \lstinline|<code>...</code>| to extract code snippets from Posts.

\subsection{Tokenization}
\label{subsec:token}
Tokenization is the process of transforming a file into a “bag of words”. Tokenization involves removing comments, spaces, tabs and other special characters, identifying each individual word (token), and counting their frequency. 

Consider the following Python block below:

\begin{lstlisting}[language=Python, label={lst:token-example},caption={Github block tokenization}]
def func1(a, b, c): # example block
  if a>b: # condition
    return c
  else:
    return 0
\end{lstlisting}

During tokenization, tokens in the block are identified and their occurrences are counted. 
The result after tokenizing the block in \ref{lst:token-example} is:

\begin{lstlisting}[numbers=none]
[(def, 1), (func1, 1), (a, 2), (b, 2), (c, 2),
(if, 1), (return, 2), (else, 1), (0, 1)]
\end{lstlisting}

\noindent
where the token \lstinline|def| and \lstinline|func1| appear once, the tokens \lstinline|a|, \lstinline|b| and \lstinline|c| appear twice, and so on.

During tokenization we also capture facts about blocks, specifically:
(1) block hash: the MD5 hash of the entire string that composes the
block; (2) token hash: the MD5 hash of the string that constitutes the
tokenized block; (3) number of lines (4) number of lines of code: LOC
(no blanks); (5) number of lines of source code: SLOC (no comments);
(7) number of tokens; (8) number of unique tokens. For GH blocks,
also (9) starting line; (10) ending line.\footnote{There is some
  possibility that hash collisions will provide the same hashes for
  different blocks. Through relevant in the fields of cryptography and
  cryptosecurity, this is so unlikely we simply chose to ignore this
  possibility.}

\subsection{Block-hash and token-hash duplicates}
Two types of code clones are calculated simply based on hash values originated from two sources: the blocks themselves, and their tokenized forms.

The first type of clones, calculated by the hash values of their absolute composition of blocks (including spaces, all the characters, comments and so on) are called block-hash clones. When two blocks are block-hash equal, it means they are an exact copy of each other.

The second type of clones are calculated using their tokenized forms. These clones, called token-hash clones, differ from block-hash clones because they focus on the source code constituents of the blocks. Note that block-hash clones provide a very precise relation between two blocks, but has the consequence of being extremely sensible to small, irrelevant variations between blocks since any minimal difference of spaces, tabs, indentation or comments for example will flag two blocks as not clones. Therefore, we use tokenization to eliminate these small idiosyncrasies between two blocks that are irrelevant from a semantic perspective.

\subsection{SourcererCC}
The first two levels of similarity are obtained by hash equality, being it at the block level or after its tokenization. These two levels do not reveal partial cloning, which in practice means certain scenarios where two blocks are cloned are not detected. Examples include familiar behaviors of literal copy-paste of a block, followed by a small specialization of a variable, or addition of tracing and debugging, actions through which intruders are inserted into the source code but their impact is so small that the blocks are still clones. This kind of problem if called near-miss clones in the area of code cloning.

To cover these scenarios, we use the tool SourcererCC \cite{Sajnani:2016}, which has the capability of detecting relative similarities of two pieces of source code given a certain threshold. SourcererCC is a token-based clone detector, it can detect three types of clones.
It also exploits an index to achieve scalability to large repositories using a standard
workstation.

By evaluating the scalability, execution time, recall and precision of SourcererCC, and comparing it to four publicly available and state-of-the-art tools, SourcererCC has been shown to have both high recall and precision, and is able to scale to a large repository using a standard workstation. All of the above make SourcererCC a good candidate for this study. We used the default settings of SourcererCC, i.e., each clone pair has more than 80\% of similarity.

\section{Dataset}
\label{sec:dataset}

We downloaded the Github (GH) Python projects by using the metadata
provided by GHTorrent
\cite{GHTorrent, Gousios:2012, Gousios:2013}. GHTorrent is a scalable, offline
mirror of data offered through the Github REST API, available to
the research community as a service. It provides access to all the
meta-data from GitHub, such as number of stars or commiters, main
languages, time points relevant to the projects and so on.

For this work, we downloaded 909k Python non-fork repositories based
on the GHTorrent's metadata available on November 2016. Filtering
non-fork projects is an important constrain because through this
mechanism information is necessarily cloned (direct replication is in
the nature of forking a project) and therefore would skew the results.

Table \ref{table:github-dataset} shows information regarding the
entire corpus of Python projects that were used in this study. The gap
between the projects that were downloaded and analyzed represents
residual problems on accessing the downloaded information (typically
corrupt zip archives, but also data on GHTorrent that was not
up-to-date). The gap between analyzed and parsed files represents
residual problems on parsing (for some reasons, Python's AST
module~\cite{AST}
could not process them); only the latter, the parsed files, contribute
to this study.


\begin{table}[]
\centering
\caption{Github Dataset}
\label{table:github-dataset}
\begin{tabular}{| l || r |} \hline
\# projects (total) & 2,340,845 \\ 
\# projects (non-fork) & 1,096,246 \\
\# projects (downloaded) & 1,096,246 \\ 
{\bf \# projects (analyzed)} & 909,288 \\
{\bf \# files (analyzed)} & 31,609,117 \\
{\bf \# parsable files (analyzed)} & 30,986,363 \\ \hline
{\bf \# parsable blocks (analyzed)} & 290,742,628 \\ \hline
\end{tabular}
\end{table}

\begin{figure*}[]
\centering{
\includegraphics[width=6.5cm]{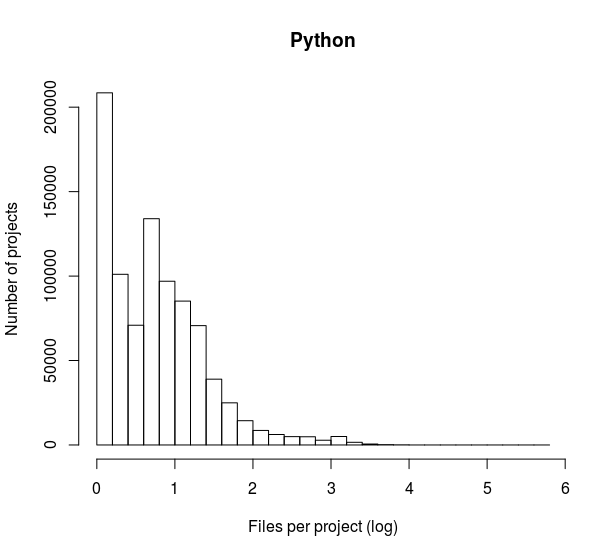}
\includegraphics[width=6.5cm]{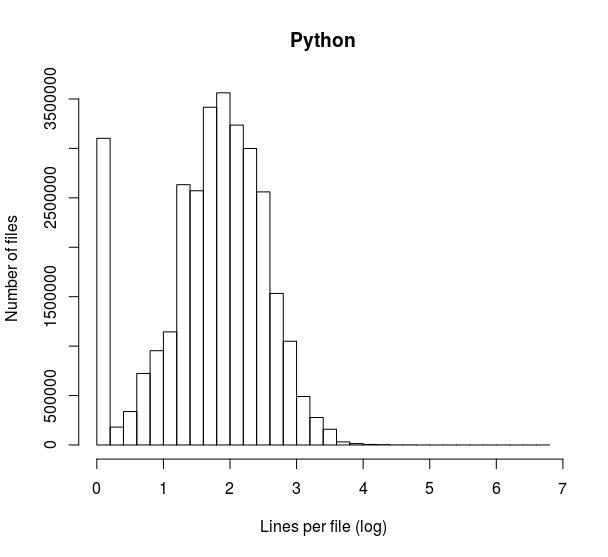}
\includegraphics[width=6.5cm]{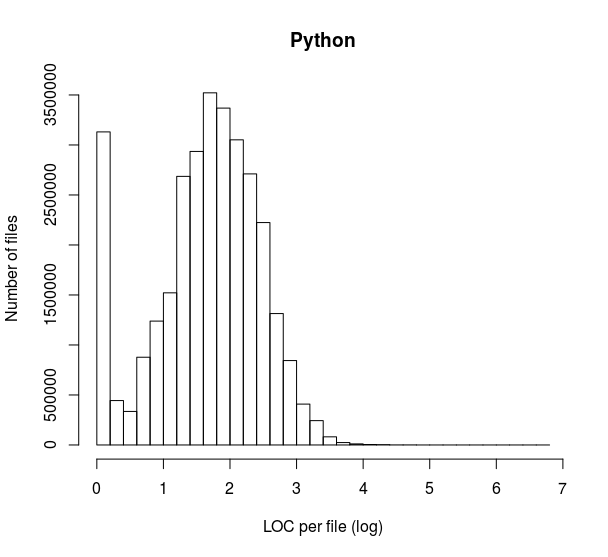}
\includegraphics[width=6.5cm]{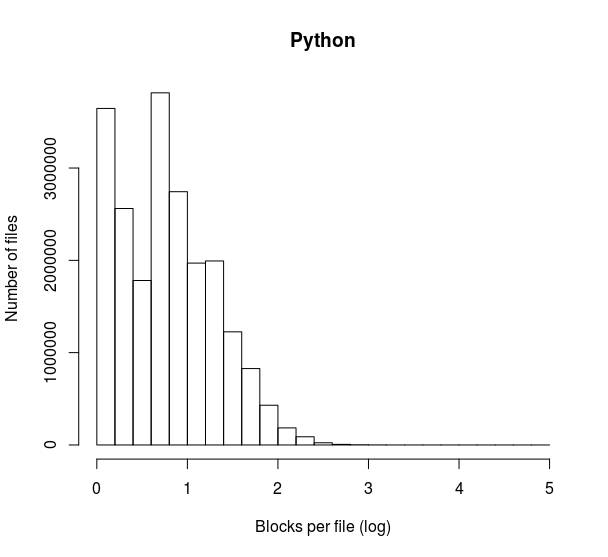}
}
\caption{Python GitHub projects. LOC means Lines Of source Code, and is calculated after removing empty lines.}
\label{fig:per-file}
\end{figure*}

Figure \ref{fig:per-file} provides information regarding basic properties of the corpus of Python projects (note the first histogram is the only one demonstrating a 'per-project' property, the others provide file's properties; and that the scale is logarithmic). 

Stack Overflow (SO) has two sources of information (two type of Posts,
from now on), typical of community-based online Q{\&}A websites: one
is the Question, and the other the Answers. All snippets were
extracted from the dump available at the Stack Exchange data dump
site~\cite{SOdump}. 

A final note: we removed single-line Python snippets because these
contain so little information that they are hardly representative. They
typically exist in the context of larger snippets for which the users
provide small comments, making them decontextualized in isolation.

\begin{figure}[]
\centering{
\includegraphics[width=6.5cm]{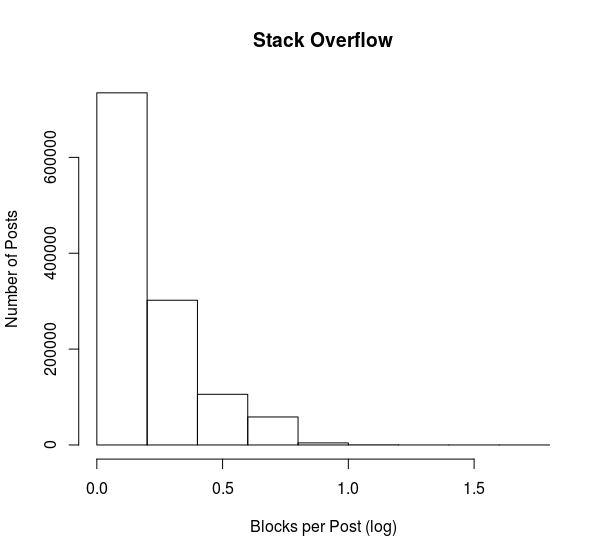}
}
\caption{Blocks per Post.}
\label{fig:blocks-per-post-SO}
\end{figure}

Table \ref{table:so-dataset} shows the total number of posts (questions and answers), number of Python blocks, and the number of multiple-line Python blocks on SO. Figure \ref{fig:blocks-per-post-SO} shows the number of blocks per post.

\begin{table}[]
\centering
\caption{Stack Overflow Dataset}
\label{table:so-dataset}
\begin{tabular}{| l || r |} \hline
\# posts (total) & 33,566,855 \\ 
{\bf \# posts (Python)} & 5,358,645 \\
{\bf \# blocks (Multiline)} & 1,954,025 \\ \hline
\end{tabular}
\end{table}

\begin{figure*}[]
\centering{
\includegraphics[width=6.5cm]{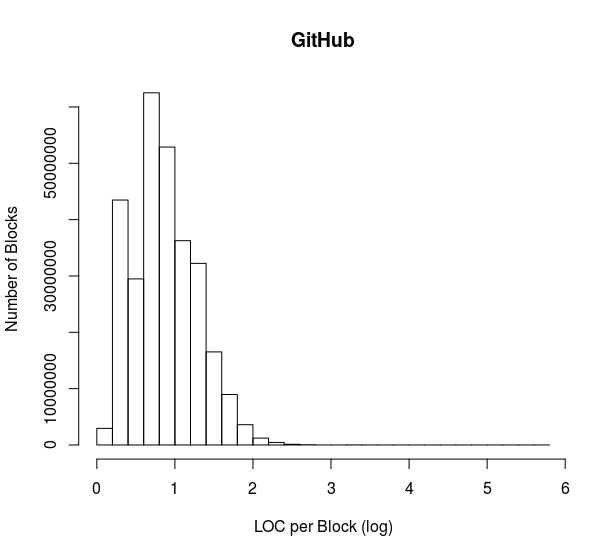}
\includegraphics[width=6.5cm]{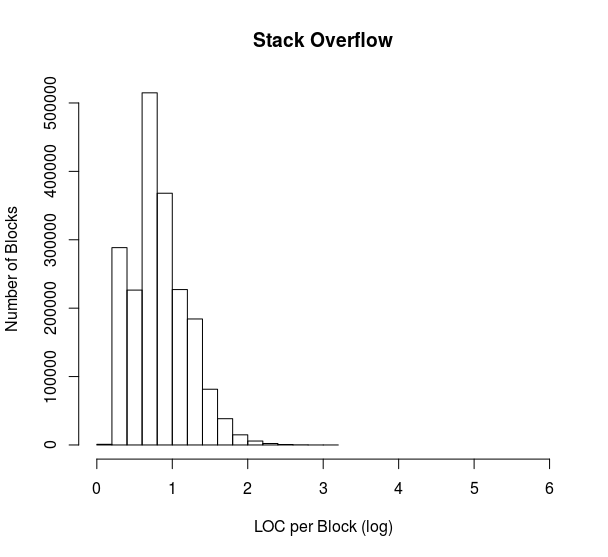}
\includegraphics[width=6.5cm]{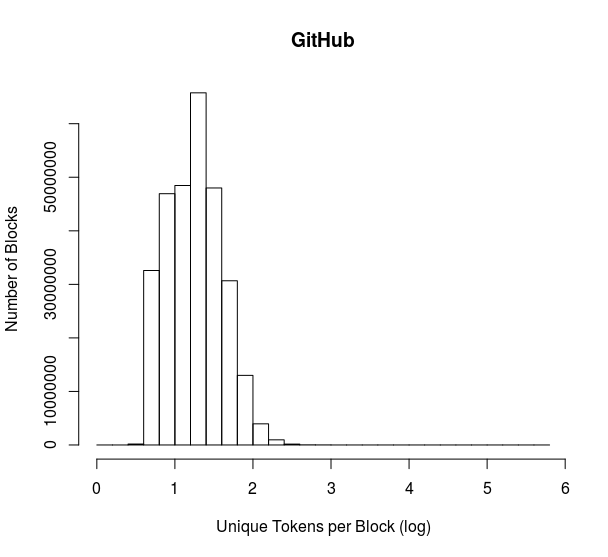}
\includegraphics[width=6.5cm]{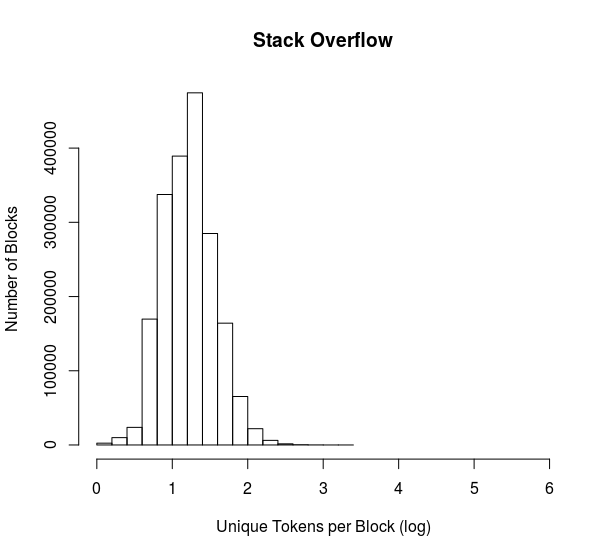}
}
\caption{Per Block distributions of LOC (top) and unique tokens (bottom), in GitHub and Stack Overflow}
\label{fig:per-block}
\end{figure*}

Figure \ref{fig:per-block} represents a comparison between blocks originated from SO and GH. On top, we can see the distribution of the number of lines of source code (total lines minus empty lines), and in the bottom we can see the distribution of unique tokens. It is interesting to observe a high degree of similarity between blocks from the two origins on the two distributions. Understanding whether this similarity is a coincidence, or the object of transport of information from one source to the other will be the object of the research presented in next Sections.

\section{Quantitative Analysis}
\label{sec:findings-quant}

In this section of we provide the values we found for code similarity between GH and SO. 

We provide three types of analysis using, first, hash values on the
blocks (block-hash), second, token hash on the blocks' source code
(token-hash) and third, partial clones using SourcererCC. This
provides different degrees of similarity for blocks: on the first
we compare for perfect equality, on the second we filter glueing
syntactic elements (spaces, tabs, terminal symbols, etc.), and on
the third we allow some divergence.

Despite focusing this work on similarities between SO and GH, we
always provide an individual analysis of each dataset. We do so to
contextualize correlations between them from the perspective of each
one individually.


\subsection{Block-hash Similarity}

The results for block-level hashing can be seen in Table
\ref{tab:bhash}. For hash analysis, we start by reducing the total
group of blocks to a distinct set of block-level hashes. This set,
shown on the second row of the table, represents the number of
distinct pieces of code on the datasets. For GH, out of the 290M
blocks there are only 40M distinct hashes, meaning that block-level
code duplication is intense: 86\% of blocks have
the same exact code as the other 14\%. This large amount of code
duplication in open source project repositories has been observed
before. For SO, the numbers are considerably smaller, with an almost
absence of block duplication; only 1.3\% of the blocks have the same
code as the other 98.7\%.

\begin{table}
\centering
\caption{Block-hash similarity}
\label{tab:bhash}
\begin{tabular}{| l || r | r |} \hline
			& GH & SO \\ \hline\hline
Total blocks  & 290,742,628 &1,954,025  \\ \hline 
Distinct block-hashes  & 40,098,522 & 1,929,411 \\\hline
Common distinct block-hashes & 1,566 & 1,566 \\ \hline
Common blocks & 60,962 & 2,091 \\ \hline
\end{tabular}
\end{table}
Next, we make the intersection of the distinct hashes in both
datasets, obtaining the common distinct hashes between GH and SO. That
number is shown in the third row: 1,566. This is a very small
percentage of the distinct hashes in both datasets.

Finally, in row four we count all the blocks whose hashes belong to
the common hashes. These are the blocks of code that exist in their
exact form, including formatting and whitespace, in both GH and
SO. The percentages are very small, less than 1\% in both cases.

\subsection{Token-hash Similarity}

The results for block-hash analysis are presented in
Table~\ref{tab:thash}. Not surprisingly, when formatting and
whitespace are ignored, the code duplication in each dataset increases
slightly, i.e. the number of distinct token hashes is smaller than the
number of distinct block hashes (compare second rows of Tables
\ref{tab:bhash} and \ref{tab:thash}).

\begin{table}
\centering
\caption{Token-hash similarity}
\label{tab:thash}
\begin{tabular}{| l || r | r |} \hline
			& GH & SO \\ \hline\hline
Total \# blocks  & 290,742,628 &1,954,025  \\ \hline 
Distinct token-hashes  & 35,894,897 & 1,890,565 \\\hline
Common distinct token-hashes & 9,044 & 9,044 \\ \hline
Common blocks & 3,839,019 & 13,747 \\ \hline
\end{tabular}
\end{table}

For the same reason, the common distinct token hashes between GH and
SO is considerably larger than the common distinct block hashes
(compare third rows of Tables \ref{tab:bhash} and
\ref{tab:thash}). But the percentage of distinct hashes that are
common to both datasets is still very small.

Interestingly, the number of blocks in GH whose token hashes are in
the common set is above 1\% (see row four). While small, it is
remarkable that so many Python functions in GH projects, almost
4M, have the exact same tokens as snippets of Python code found in SO.





\subsection{SCC Similarity}

The analysis in this subsection is slighly different than in the
previous two: we narrow the analysis only to the universe of blocks
that have distinct token hashes, those counted in the second row of
Table~\ref{tab:thash}. The rationale is that two files with the same
token-hashes will be detected as clones by SCC, and
therefore it suffices to process only one representative of each group of
blocks with the same token hash.

\begin{table}
\centering
\caption{SCC Similarity}
\label{tab:scc}
\begin{tabular}{| l || r | r |} \hline
			& GH & SO \\ \hline\hline
Distinct token hashes  & 35,894,897 &1,890,565  \\ \hline 
SCC-dup  & 13,363,759 & 297,554 \\\hline
Common  &  405,393 & 35,098 \\ \hline
\end{tabular}
\end{table}

The results are presented in Table~\ref{tab:scc}. The second row,
SCC-dup, shows the number of blocks in each dataset that have at least
one similar block in the same dataset -- only within the
universe of distinct token hashes. The amount of near-duplication is
considerably high in GH (roughly, 37\%), but less in SO (roughly
16\%).

The third row shows the number of blocks that are similar between
datasets -- again, only within the universe of distinct token
hashes. More than 405k (1.1\%) of the blocks
in GH are similar to blocks in SO, and 2\% of blocks in
SO are similar to blocks in GH. This means that 35,098 distinct
blocks found in SO can be found in very similar form in GH. The number is considerably larger than the common distinct token-hashes in Table~\ref{tab:thash}.

\section{Qualitative Analysis}
\label{sec:findings-qual}

To understand the nature of blocks that can be found in both SO
and GH, we made a qualitative analysis on the duplicated blocks. This analysis was made
in two steps. We first looked at subsets of all of them,
looking for patterns. One strong pattern emerged: the majority of
blocks that are duplicated -- both within datasets as well as between
them -- are very small, typically a couple of lines of code. These
also tend to be non-descriptive, with very generic code (e.g. trivial
\lstinline|__init__| methods). Having observed this, we then moved to a second
stage of analysis, where we looked only at larger functions. The
number of these blocks is much smaller, but they are more interesting.
This section describes our qualitative analysis.

\subsection{Step 1: Duplicated Blocks}

We looked the top 10 most duplicated code blocks based on their
block-hash, token-hash, and SourcererCC reported clones. We did this
analyses for intra-GH and intra-SO block clones.
Further, to understand the kind of code blocks which are common across
GitHub and SO, we looked at the 10 code blocks which are present in
both GitHub and SO, and are duplicated the most in GitHub and
similarly the top 10 code blocks which are duplicated the most in
SO. The duplicated code blocks were selected based on block-hash,
token-hash, and SourcererCC reported clones.

\subsubsection{Block-Hash Duplicates} \mbox{}\

\textbf{Intra-GH}: All of the top 10 duplicated code-blocks had 2
lines of code. Four of these methods can be traced back to
\lstinline|cp037.py| file, located at
\url{https://github.com/python-git/python/blob/master/Lib/encodings/cp037.py}. The
file gets generated from
\lstinline|'MAPPINGS/VENDORS/MICSFT/EBCDIC/CP037.TXT'| with
\lstinline|gencodec.py| as mentioned in the file level comment inside
the file. There are many such files, each for a different encoding
\lstinline|cp1253|, \lstinline|cp1026|, \lstinline|cp1140|, and so
on. Further, we found many instances where these files are present in
other GitHub projects. Each of these files contains the generic
methods to encode and decode the input string, as shown in the
Listing~\ref{lst:codec_example} below.

\begin{lstlisting}[language=Python, label={lst:codec_example},caption={Most duplicated code block based on Block-Hash},frame=tlrb]
class Codec(codecs.Codec):

  def encode(self,input,errors='strict'):
    return codecs.charmap_encode(input,errors,encoding_table)

  def decode(self,input,errors='strict'):
    return codecs.charmap_decode(input,errors,decoding_table)
\end{lstlisting}

Other most duplicated code blocks are private methods \lstinline|__iter__|, \lstinline|__enter__|, \lstinline|__ne__|, and \lstinline|__init__|, with just one statement, as shown in Listing~\ref{lst:private_example}. 
\begin{lstlisting}[language=Python, label={lst:private_example},caption={Example of one highly duplicated private method},frame=tlrb]
def __iter__(self):
  return self
\end{lstlisting}

\textbf{Intra-SO}: Like GH, the top 10 most duplicated code blocks on SO have two lines of code. Listing~\ref{lst:SO_example} shows the most duplicated ones. The first code block (top), shows Python idiom for the main entry point in a module. The second code block from the top, prints out all directories which are on the python's path. This is usually done to fix issues related to the import of third party libraries. The bottom two blocks, also very common on SO, do not have any Python specific code, and are used to present an example output of some Python code. 

\begin{lstlisting}[language=Python, label={lst:SO_example},caption={Most duplicated code blocks on SO based on Block-Hash},frame=tlrb]
if __name__ == '__main__':
  main()
\end{lstlisting}
\begin{lstlisting}[language=Python,frame=tlrb]
import sys
print sys.path

\end{lstlisting}
\begin{lstlisting}[language=Python,frame=tlrb]
True
False
\end{lstlisting}
\begin{lstlisting}[language=Python,frame=tlrb]
1
2
\end{lstlisting}

\textbf{Most duplicated blocks in GH that are also present in SO}: Listing~\ref{lst:MGH_SO} shows two of the most duplicated blocks in GH that are also present in SO. We found the code block for \lstinline|session()| on SO, where it is mentioned that this code blocks was copied from the sessions module under the requests library. We found that a lot of projects on GH use this library, where they copy the entire source code. \lstinline|__iter__| is a very common private function used to make a class iterable, and hence this code block is also duplicated a lot. On SO we found a post where this code block was used as an example to demonstrate how to make a class iterable.

\begin{lstlisting}[language=Python, label={lst:MGH_SO},caption={Most duplicated code blocks on GH, which are also present in SO based on Block-Hash},frame=tlrb]
def session():
  """Returns a :class:`Session` for context-management."""

  return Session()
\end{lstlisting}
\begin{lstlisting}[language=Python,frame=tlrb]
def __iter__(self):
  return self

\end{lstlisting}

We also found some code blocks which are related to \textit{Django}, a python web framework. These code blocks are not intentionally copied, and become a part of the projects that are using \textit{Django}. Listing~\ref{lst:MGH_Django_SO} shows an example code block. On inspection we found that this code block was copied to SO from GH, to show the code in Django which is responsible for creating an anonymous user.

\begin{lstlisting}[language=Python, label={lst:MGH_Django_SO},caption={Most duplicated code blocks on GH, which are also present in SO based on Block-Hash},frame=tlrb]
def get_user(request):
  from django.contrib.auth.models import AnonymousUser
  try:
    user_id = request.session[SESSION_KEY]
    backend_path = request.session[BACKEND_SESSION_KEY]
    backend = load_backend(backend_path)
    user = backend.get_user(user_id) or AnonymousUser()
  except KeyError:
    user = AnonymousUser()
  return user
\end{lstlisting}

An observation common to most of these code blocks is that these blocks get duplicated in GH not because developers are interested in a particular code block, but because they are interested in the entire module like modules from \textit{requests} library, or because they are using a framework which adds the source files into the projects. 
On SO, users are more interested in explaining a particular behavior or seeking some explanation about code blocks. We observed such scenarios where users have used a code block from GH and have also pasted the link of the source file in GH. 

\textbf{Most duplicated blocks in SO that are also in GH}: Interestingly, 8 out of the top 10 code blocks come from \lstinline|itertools| \url{https://docs.python.org/2/library/itertools.html#itertool-functions}. To understand the origin of these code blocks on SO, we looked at the two most duplicated ones, shown in Listing~\ref{lst:MSO_GH}. On SO, we found 28 instances of the block on top, \lstinline|any()| and 24 instance of the one in bottom, \lstinline|grouper()|. On SO, we looked at 5 random instances of \lstinline|any()| and found that this code block was copied from \url{https://docs.python.org/2/library/functions.html#any} and not from GH. We could link the origin based on the comments written on the SO posts. On GH we found some projects which have \lstinline|any.py| module implementing the exact code block. We also found modules on GH which implement code blocks similar to \lstinline|any| like \lstinline|all|, \lstinline|enumerate|. Some of these modules come from projects where it was quite evident that the user has copied code into their project. For example, a project where a duplicate of \lstinline|any()| function found, mentions in its \lstinline|README.md|: \textit{I want to collect something that I think it's interesting. Maybe some code snippet I think it's excellent cool}.

We made a similar observation when we looked at the origin of \lstinline|grouper()|. In many instances on SO, the code block was copied from the python docs. On GH, we found one instance of this code block at \url{https://github.com/hbradlow/dynamic_path/blob/master/path/utils.py}. We also observed a comment in the same file with a url to a SO post. On further inspection we found that the most of the code in the module was copied from the SO post.

\begin{lstlisting}[language=Python, label={lst:MSO_GH},caption={Most duplicated code blocks on SO, which are also present in GH based on Block-Hash},frame=tlrb]
def any(iterable):
  for element in iterable:
    if element:
      return True
  return False
\end{lstlisting}
\begin{lstlisting}[language=Python,frame=tlrb]
def grouper(n, iterable, fillvalue=None):
  "grouper(3, 'ABCDEFG', 'x') --> ABC DEF Gxx"
  args = [iter(iterable)] * n
  return izip_longest(fillvalue=fillvalue, *args)
\end{lstlisting}

\subsubsection{Token-Hash Duplicates} \mbox{}\

\textbf{Intra-GH}: To analyze Token-Hash duplicates, we followed a process similar to what we used for analyzing Block-Hash duplicates. The observations are very similar to those made in the Block-Hash duplicates section. Most duplicated code blocks are \lstinline|encode|, \lstinline|decode|, \lstinline|__iter__|, \lstinline|__enter__|, \lstinline|__ne__|, and \lstinline|__init__|, as shown in the Listings~\ref{lst:codec_example} and~\ref{lst:private_example}.
 
\textbf{Intra-SO}: We found that the code blocks that resulted into 0 tokens were reported as the most duplicated code blocks. These are the blocks where all statements are commented, for example consider a code block shown in Listing~\ref{lst:EMPTY}. This blocks will generate 0 tokens as comments are ignored during tokenization. The token hash of all such blocks will be computed on an empty string, resulting into same token-hash.

\begin{lstlisting}[language=Python, label={lst:EMPTY},caption={Example of 0 token code block},frame=tlrb]
#define private public
#include <module>
\end{lstlisting}

Listing~\ref{lst:SO_example_Thash} shows examples of duplicated code blocks on SO. The first code block (top), shows an example of how to instantiate three different list objects. The SO post for this code block is full of similar examples. 
 
\begin{lstlisting}[language=Python, label={lst:SO_example_Thash},caption={Most duplicated code blocks on SO, based on Token-Hash},frame=tlrb]
a = []
b = []
c = []
\end{lstlisting}
\begin{lstlisting}[language=Python,frame=tlrb]
1 2 3
4 5 6
7 8 9
\end{lstlisting}

The second code block from the top, shows a representation of a two dimensional list. SO, has many such blocks, where users have used this representation to explain the desired output of their code. 5 out of top 10 duplicated blocks on SO are about lists of numbers. We also observed many code blocks similar to shown in Listing~\ref{lst:SO_example}.

\textbf{Most duplicated blocks in GH that are also present in SO}: The results found are mostly shared code represents simple two liners and are very trivial, such as \lstinline|__ne__| or \lstinline|__str__|.
 
\textbf{Most duplicated blocks in SO that are also in GH}: Similarly to the results on the opposite direction, the blocks we found were of a small dimension and were characterized by trivial information.

\textbf{Token hash vs Block hash} We ignored the output explanation blocks, and dig into the reason for the code block pairs being caught as duplicates for token-hash level instead of block-hash level. We found that most of the pairs were only different in spaces. Some are token-hash duplicates because of the difference in the syntax of Python 2 and Python 3, for example in the \lstinline|print| function. A few are token-hash duplicates because some parameter or variable is set to be am empty list, which results in differences in special characters, for example in the pair in List \ref{lst:thash-small}.

\begin{lstlisting}[language=Python, label={lst:thash-small},caption={Example for difference in token-hash duplicates},frame=tlrb]
def __init__(self, connection):
	self.connection = connection
\end{lstlisting}
\begin{lstlisting}[language=Python,frame=tlrb]
def __init__(self, connection=[]):
        self.connection = connection
\end{lstlisting}

\subsubsection{SourcererCC Duplicates}\mbox{}\

The qualitative analysis of block-hash duplicates and token-hash duplicates hints at exact copy-paste inside and between GH and SO. However, from the SCC results in the quantitative analysis, we learned that there are many cases where programmers make adaptations to the codes during copy-pasting. Therefore, here we want to see how people change their code when inside and between GH and SO.

\textbf{Intra-GH}: All of the top 10 most duplicated blocks we found inside GH are from the same file, located at \url{https://github.com/lufo816/WeiXinCookbook/blob/master/urlHandler.py}. This file has 80,452 clones similar to the block on Listing \ref{lst:urlHandler_example}.

\begin{lstlisting}[language=Python, label={lst:urlHandler_example},caption={Most duplicated code block intra GH from SCC},frame=tlrb]
def GET(self):
    return render.caipu1()
\end{lstlisting}

The only difference on these blocks is the number in the function, ranging from 1 to 80,452. SCC will take every block as a clone for all other blocks in this same file, so they become the most duplicated blocks.

\textbf{Intra-SO}: Here, we found that 5 of the 10 examples are python error message of \lstinline|ImportError|, similar to the one on Listing \ref{lst:ghfjfj}, with the difference in module name or line number. There were 4 blocks that are standard settings for \textit{Django} and the remaining one is a list of numbers representing an output, similar to what we have seen above.

\begin{lstlisting}[language=Python, label={lst:ghfjfj},caption={A common error message},frame=tlrb]
Traceback (most recent call last):
  File "<stdin>", line 1, in <module>
ImportError: No module named MySQLdb
\end{lstlisting}

\textbf{Most duplicated blocks in GH that are also present in SO}: Overall, 7 out of 10 of the blocks we analyzed can be traced to modules from libraries like \lstinline|requests| and \lstinline|pip|, such as the examples of Listing \ref{lst:asdasdadsfasf}.

\begin{lstlisting}[language=Python, label={lst:asdasdadsfasf},caption={Most duplicated code blocks on SO, which are also present in GH based on the results from SCC},frame=tlrb]
    def __ne__(self, other):
        return not self.__eq__(other)
\end{lstlisting}

\begin{lstlisting}[language=Python,frame=tlrb]
    def __init__(self, username, password):
        self.username = username
        self.password = password
\end{lstlisting}
 
\textbf{Most duplicated blocks in SO that are also present in GH}: For the top 10 most duplicated blocks in SO that are also present in GH, there are actually only three kinds of blocks as shown in \ref{lst:inter-SO}. The first is a standard initial function for a class; the second is a standard function definition with parameters, the third is a function that raise \lstinline|NotImplementedError|. The first group contains 5 pairs, the second contains 4 pairs, and the third only has 1 pair.

\begin{lstlisting}[language=Python, label={lst:inter-SO},caption={Most duplicated code blocks on SO, which are also present in GH based on SCC},frame=tlrb]
def __init__(self):
    self.locList = []
\end{lstlisting}
\begin{lstlisting}[language=Python,frame=tlrb]
def some_function(*args, **kwargs):
    pass
\end{lstlisting}
\begin{lstlisting}[language=Python,frame=tlrb]
def number_of_edges(self):
    raise(NotImplementedError)
\end{lstlisting}

When observing the SCC clone for each block, we found that for in group 1, all pairs contain the tokens \lstinline|def|, \lstinline|raise|, and \lstinline|self|, and the only difference is the adaptation to specific variables. For group 2, all clone pairs contain tokens \lstinline|def|, \lstinline|*args|, \lstinline|**kwargs|, \lstinline|pass|, the only difference is the function name. For the block in group 3, its clone pair have the same tokens \lstinline|def|, \lstinline|self|, \lstinline|raise| and \lstinline|NotImplementedError| and the only difference is the function name.

\textbf{SCC vs Token hash} From the observations above, we can see that the reason for these pairs being duplicates in SCC level instead of token-hash level is changes of function names, parameters, or variables.

\subsection{Step 2: Large Blocks Present in GH and SO}

To further understanding the correlation and copy-paste behavior between GH and SO, we set a threshold to the number of unique tokens in a block to get larger blocks for observation.

\subsubsection{Block-Hash Duplicates} \mbox{}\

We set the threshold of unique tokens of each block to be equal or larger than 30 tokens in order to filter out meaningless small blocks. This filtering left us with 104 common block hashes between GH and SO, from the original 1,566 common block hashes. We sampled 10 block hashes out of these 104 and traced one sample pair of GH and SO blocks for each common block hash.

From the 10 pairs we got, 4 of the GH blocks explicitly stated in the comments that the code was borrowed from SO, and also gives the SO post link corresponding to the block. The SO post links were exactly the same as we paired for the GH block. This is a very clear evidence source code has been flowing from SO to GH.

In two pairs, GH and SO blocks are coming from the same third-party source. In another three pairs, the SO post stated that the code was copied from a third-party source, but there's no explicit clue of where the GH block comes from, although there was only one commit on the file and no changes before and after, which may indicate the code was copied from other sources too.

\subsubsection{Token-Hash Duplicates} \mbox{}\ 

The number of unique tokens is set to be equal or larger than 35 for token hash duplicates. We have 915 common token hashes between GH and SO after the filtering. For large token-hash duplicates, we observed a clear case of copy-paste from GH to SO, where the author of the code on GH used his own code as an example to demonstrate aspects of Python's \lstinline|func_code| attribute.

Another relevant example is where a closer inspection of the comments on SO pointed directly the original website from where these blocks were copied, which happens to be the now defunct Google code. There are two clear indications of the transfer of knowledge from one source to the other.

Then we furthered observed the reason for the pairs being caught as duplicates only by token hash instead of block hash. Although token hash will leaving out all the comments, spaces, special characters, nine out of the ten sampled pairs only different in spaces, and all contents are kept as-is, including comments; only one pair is different in missing one line of comments. It means that during the process of copy-pasting large blocks of code, either between GH and SO or from other sources, programmers tend to preserve everything instead of dropping or changing any of them. This may because on one hand, the large blocks are a complete implementation of some functionality, and plugging them as-is is sufficient for programmers' needs and no changes needed; on the other hand, copy-paste is also a process of learning, and the comments help the learner understand what the code is about, so there's no point of deleting them intentionally.


\subsubsection{SourcererCC Duplicates} \mbox{}\

The number of unique tokens is set to be equal or larger
than 35 for SCC duplicates. We have 4699 distinct token
hashes in SO that can be found very similar form in GH.

Using SCC we found many cases where code blocks on SO were similar to that on GH. On observing the blocks manually, it was hard to find clues that point at the directional of information exchange. In some cases it was obvious that deliberate copy-paste has resulted into code duplication, but we cannot say for sure whether the code was copied from GH to SO, SO to GH or from a third party website to GH or SO. 

SCC marked these pairs as 80\% similarity in tokens. We observed that the differences between them came from variables, function identifiers,  \lstinline|if| conditions, or \lstinline|class| definition. In other words, when copy-pasting codes, programmers will adjust the variables, switch function names or parameters, change, add, or delete \lstinline|if| conditions, or add or delete \lstinline|class| definition to match their particular needs. 

\section{Related Work}
\label{sec:related-work}
This paper involves different aspects of study, first, it focuses on the code itself of SO; second, it discovers the relationship between SO and GH; third, it investigates the large-scale code duplication detection in block-level, which includes uniqueness of source code. The related work come with these angles.

Wong et al. \cite{Wong:2013} devised a tool that automatically generates comments for software projects by searching for accompanying comments to SO code that are similar to the project code. They did so by relying on clone detection. 
This work is very similar to Ponzanelli et al. \cite{Ponzanelli:2013} \cite{Ponzanelli:MSR2014} \cite{Ponzanelli:ICSME2014} in terms of the approach adopted. Both mine for SO code snippets that are clones to a snippet in the client system, but Ponzanelli et al.'s goal was to integrate SO into an Integrated Development Environment (IDE) and seamlessly obtain code prompts from SO when coding. In another work from Ponzanelli et al., they presented an Eclipse plugin, Seahawk, that also integrates SO within the IDE. It can add support to code by linking programming tools with SO search results.

There are two studies about assessing the usablity of code in SO. Nasehi et al. \cite{Nasehi:2012} engaged in finding the characteristics of a good example. They adopted a holistic approach and analyzed the characteristics of high voted answers and low voted answers. They enlisted traits by analyzing both the code and the contextual information: the number of code blocks used, the conciseness of the code, the presence of links to other resources, the presence of alternate solutions, code comments, etc. 

Yang \cite{Yang:2016} assessed the usability of SO snippet with a different criteria. They  define usablity based on the standard steps of parsing, compiling and running the source code, which indicates that the effort that would be required to use the snippet as-is. A total of 3M code snippets are analyzed across four languages: C\#, Java, JavaScript, and Python. Python and JavaScript proved to be the languages for which the most code snippets are usable. Conversely, Java and C\# proved to be the languages with the lowest usability rate.

Vasilescu, et al. \cite{Vasilescu:2013} investigated the interplay between SO activities and the development process, reflected by code changes committed to the largest social coding repository, GH. They found that active GH committers ask fewer questions and provide more answers than others, and active SO askers distribute their work in a less uniform way than developers that do not ask questions.

An \textit{et al.}~\cite{An:2017} aims to raise the awareness of the software engineering community about potential unethical code reuse activities taking
place on Q\&A websites like SO. They conducted a case study with 399 Android apps, to investigate whether developers respect license terms when reusing code from SO posts (and the other way around). From the 232
code snippets in 62 Android apps that were potentially reused from SO, and the 1,226 SO posts containing code examples that are clones of code released in 68 Android apps,  they observed 1,279 cases of potential license violations (related to code posting to SO or code reuse from SO).

Some previous work has been done on code clone detection in block-level. Roy and Cordy \cite{Roy:2010} analyzed clones in twenty open source C, Java and C\# systems, using the NiCad block-level clone detector. They found that on average 15\% of the files in the C systems, 46\% of the files in the Java systems and 29\% of files in the C\# systems are associated with exact (block-level) clones. Heinemann et al.\cite{Heinemann:2011} computed type-2 block-level clones  between selected 22 commonly reused Java frameworks (e.g. Eclipse and Apache)  and 20 open source Java projects. They didn’t find any clones for 11 of the 20 study objects. For 5 projeccts, they found cloning to be below 1\% and for the remaining 4 projects, they found in the range of 7\% to 10\% cloning.

Gabel \textit{et al.}~\cite{Gabel:2010} presented the results of the first study of \textit{uniqueness of source code}. They gave \textit{uniqueness} of a unit of source code a precise measure: synctactic redundancy. They wanted to figure out at what levels of granularity is software unique, and at a given level of granularity, how unique is software. We compute syntactic redundancy for 30 assorted SourceForge projects
and 6,000 other projects. The results revealed a general lack of uniqueness in software at levels of granularity equivalent to approximately one to seven lines
of source code. This phenomenon appears to be pervasive, crossing
both project and programming language boundaries.

Hindle \textit{et al.}~\cite{Hindle:2012} pointed out like natural language, software is also likely to be repetitive and predictable. Using n-gram model, they provided empirical
evidence to support that code can be usefully modeled by statistical language models and such models can be leveraged to support software engineers.
They showed that code is also very repetitive, and in fact even more
so than natural languages.

\section{Conclusion}
\label{sec:conclusion}

Stack Overflow, a popular Q\&A site, has become one of the major
Internet hubs where programmers can find all sorts of information
related to simple, but concrete programming problems. We wanted to
find out the extent to which the code snippets in SO find their way to
open source projects. For this study, we focused on programs written
in Python. As datasets, we took the collection of 909k non-forked
Python projects hosted in Github, as well as the SO dump provided by
Stack Exchange. We extracted all the multi-line Python code snippets
from SO, and we parsed all the Python projects, breaking them into
functions. We then cross referenced the SO snippets with these
functions, using three measures of similarity: exact match, match on
the tokens and near-duplication as detected by a code clone detector
tool.

Our quantitative analysis shows that exact duplication between SO and
GH exists, but is rare, much less than 1\%. Token-level duplication is
more common, with almost 4M blocks in GH being similar to SO
snippets. In terms of percentage, this is still
small. Near-duplication shows 405k distinct blocks (1.1\%) in GH being similar to SO and 35k (2\%) SO distinct blocks having near duplicates in GH. Although the percentages are not very large, the numbers are in thousands.

Upon careful qualitative analysis, we observed that the vast
majority of these duplicates are very small, typically 2 lines of code
and just a few tokens. Moreover, they tend to be non-descriptive,
meaning that they are too generic to trace. Because they are generic
and small, likely they didn't flow from SO to GH or vice versa. We
then focused out attention to the fewer blocks that are not so
small. For these, we found evidence that there is, indeed, flow from
SO to GH, in some cases that flow being explicitly stated in
comments. While there is a lot less of these, their number is still in
the thousands. 

The importance of this work is twofold. First, it gives empirical
evidence of the phenomenon of copy-and-paste from SO,
something that is widely accepted to be true, but hasn't been
studied. Second, the non-trivial SO snippets that can be found in real
code in GH could be used as the basis for novel search engines for
program synthesis and repair that integrate with the rich natural
language descriptions found in SO. We found that there are 5,718 large blocks 
with distinct hashes in SO that can be found very similar form in GH. These
large distinct blocks can be made good use of in the future work.

Enriched by natural language contexts surrounding the code snippets, SO can help to retrieve code snippets by matches on the non-coding information. Moreover, it can potentialy be used as a knowledge base for tools that automatically combine snippets of code in order to obtain more complex behavior.
The viability of using SO in program synthesis lies, first of all, on the existence of good snippets and evidence that they exist in real code, which is shown in this paper.

\bibliographystyle{IEEEtran}
\bibliography{msr}

\end{document}